\documentstyle[12pt]{article}
\makeatletter
\def\@maketitle{\newpage
 \null
 {\normalsize \tt \begin{flushright}
  \begin{tabular}[t]{l} \@date
  \end{tabular}
 \end{flushright}}
 \begin{center}
 \vskip 2em
 {\LARGE \@title \par} \vskip 1.5em {\large \lineskip .5em
 \begin{tabular}[t]{c}\@author
 \end{tabular}\par}
 \end{center}
 \par
 \vskip 1.5em}

\newif\if@defeqnsw \@defeqnswtrue

\def\eqnarray{\stepcounter{equation}\let\@currentlabel=\theequation
\if@defeqnsw\global\@eqnswtrue\else\global\@eqnswfalse\fi
\tabskip\@centering\let\\=\@eqncr
$$\halign to \displaywidth\bgroup\hfil\global\@eqcnt\z@
  $\displaystyle\tabskip\z@{##}$&\global\@eqcnt\@ne
  \hfil$\displaystyle{{}##{}}$\hfil
  &\global\@eqcnt\tw@ $\displaystyle{##}$\hfil
  \tabskip\@centering&\llap{##}\tabskip\z@\cr}

\@namedef{eqnarray*}{\@defeqnswfalse\global\@eqnswfalse\eqnarray}
\@namedef{endeqnarray*}{\endeqnarray}

\def\yesnumber{\global\@eqnswtrue}

\def\@@eqncr{\let\@tempa\relax\global\advance\@eqcnt by \@ne
    \ifcase\@eqcnt \def\@tempa{& & & &}\or \def\@tempa{& & &}\or
     \def\@tempa{& &}\or \def\@tempa{&}\else\fi
     \@tempa \if@eqnsw\@eqnnum\stepcounter{equation}\fi
     \if@defeqnsw\global\@eqnswtrue\else\global\@eqnswfalse\fi
     \global\@eqcnt\z@\cr}

\newtoks\@stequation

\def\subequations{\refstepcounter{equation}%
  \edef\@savedequation{\the\c@equation}%
  \@stequation=\expandafter{\theequation}
  \edef\@savedtheequation{\the\@stequation}
  \edef\oldtheequation{\theequation}%
  \setcounter{equation}{0}%
  \def\theequation{\oldtheequation\alph{equation}}}

\def\endsubequations{%
  \ifnum\c@equation < 2 \@warning{Only \the\c@equation\space subequation
    used in equation \@savedequation}\fi
  \setcounter{equation}{\@savedequation}%
  \@stequation=\expandafter{\@savedtheequation}%
  \edef\theequation{\the\@stequation}%
  \global\@ignoretrue}

\makeatother

\newcommand{\tr}{\mbox{tr}}

\newcommand{\EQ}[1]{Eq.(\ref{#1})}
\advance\oddsidemargin by -\textwidth
\advance\oddsidemargin by -1in
\evensidemargin=\oddsidemargin
\newcommand{\tMCpb}[1]{{\hat \alpha}_{#1\perp}}    
\newcommand{\tMCpr}[1]{{\hat \alpha}^{#1}_\perp}   
\newcommand{\tMClb}[1]{{\hat \alpha}_{#1\parallel}}
\newcommand{\tMClr}[1]{{\hat \alpha}^{#1}_\parallel}
\newcommand{\cR}{{\cal R}} 
\newcommand{\cL}{{\cal L}} 
\newcommand{\cV}{{\cal V}} 
\newcommand{\cA}{{\cal A}} 
\newcommand{\eq}[1]{(\ref{#1})}
\def\Re{\mathchar"023C{\rm e}}
\title{
  Chiral Perturbation to One Loop \\
  Including the $\rho$ Meson
}
\author{
  {\sc Masaharu Tanabashi}\thanks{
    E-mail address: {\tt tanabash@theory.kek.jp}}
 \\
  {\it National Laboratory for High Energy Physics (KEK)} \\
  {\it Tsukuba, Ibaraki 305, Japan}
}
\date{
  KEK-TH-349  \\
  KEK preprint 92-169 \\
  April, 1993
}
\begin{document}
\maketitle
\begin{abstract}
     We formulate the chiral perturbation theory at the one loop level
     in the effective lagrangian including the $\rho$ meson
     as a dynamical gauge boson of a hidden local symmetry(HLS).
     The size of radiative correction to the phenomenological parameter
     $a$ of HLS is estimated to be about $10$\%.
     The complete list of ${\cal O}(E^4)$ terms is given and the one
     loop counter terms are determined explicitly in the $N$ flavor model.
     We also obtain matching conditions to the conventional chiral
     perturbation of Gasser and Leutwyler in the chiral limit
     in a renormalization scale independent manner.
     We find that Gasser--Leutwyler's estimates for $L_{9,10}$
     are saturated by $\rho$ and its one loop contributions without
     introducing non-minimal couplings of $\pi$-$\rho$ system,
     suggesting the absence of the tree level $a_1$ meson contributions.
\end{abstract}

Chiral perturbation theory (ChPT)\cite{kn:We79b,kn:GL84}
is a framework to describe low energy nature of the spontaneous chiral
symmetry breaking in QCD\@.
It gives a systematic low energy expansion of QCD amplitudes
in terms of the number of derivatives appearing in the effective lagrangian.
Based on the well-known isomorphism between the electroweak symmetry
breaking and the spontaneous chiral symmetry breaking in QCD,
ChPT becomes very popular also in the physics of the Higgs sector.
Actually, the method of chiral lagrangian offers a perspective view in
the analysis of precision tests of the Higgs sector\cite{kn:HT90}.

The effective lagrangian (chiral lagrangian) written in terms of
the Nambu-Goldstone (NG) boson field with the lowest order of
derivatives is ${\cal O}(E^2)$ ($\equiv {\cal O}(\partial^2)$),
which reproduces results of the low energy theorems.
Taking account of the ${\cal O}(E^4)$ terms and the one loop corrections,
we can extract further information from the chiral
symmetry\cite{kn:We79b,kn:GL84} than those from the low energy theorems.

It is evident, however, that the ChPT cannot be applied at the scale of
the mass of the $\rho$ meson (the lightest non-NG boson) in QCD\@.
We need to introduce an explicit degree of freedom corresponding to
the $\rho$ meson so as to make the effective lagrangian valid at
the scale of its mass.
Bando, Kugo, Uehara, Yamawaki and Yanagida (BKUYY)\cite{kn:BKUYY}
constructed the most successful effective lagrangian of the $\pi$-$\rho$
system based on the idea of hidden local symmetry.
This lagrangian,
being identical with Weinberg's model\cite{kn:We68} in the unitary gauge,
describes successfully some phenomenological properties
in the $\pi$-$\rho$ system such as
the $\rho$ coupling universality\cite{kn:SaTXT}
(and thus, the $\rho$ dominance\cite{kn:SaTXT})
and the KSRF relation\cite{kn:KS66}
by choosing its phenomenological parameter $a=2$.
The ``vector limit'' lagrangian\cite{kn:Ge89}
can be also considered as a special case of
the model at $a=1$.

The BKUYY lagrangian has been applied to various processes at tree level.
Analyses of kaon decays in the BKUYY lagrangian have been performed
at tree level with a fixed (non-running) parameter $a$\cite{kn:Ko92}.
The effect of the $\rho$ meson on the Gasser--Leutwyler
parameters $L_i$\cite{kn:GL84} of the conventional ChPT is also
investigated\cite{kn:EGPR89}.
The lagrangian is known as the BESS model when it is applied to the Higgs
sector\cite{kn:BESS}.
Those tree level analyses, however,
leave uncertainties from the radiative corrections.
On the other hand,
the analyses at the one loop level have been dealt with a few papers:
Cveti\v{c} and K\"{o}gerler\cite{kn:CK91} calculated
divergent coefficients in the self-interactions of the gauge fields in
the BESS model.
Harada and Yamawaki\cite{kn:HY92} showed that the low energy
theorem\cite{kn:BKY85} of the hidden local symmetry (HLS)
is maintained at one loop in the Landau gauge.\footnote{
Although the low energy theorem of HLS
holds only at the off-shell of the $\rho$
meson and is not directly related to the physical amplitudes,
the phenomenological success indicates that the low energy theorem
is smoothly extrapolated to the on-shell of the $\rho$ meson.}
However, those one loop level calculations are not sufficient to
estimate the size of radiative corrections in the BKUYY model
of $\pi$-$\rho$ system.

In this paper, we address this problem by
formulating the ChPT of the BKUYY effective lagrangian\cite{kn:BKUYY}
in a systematic manner at one loop.
The complete list of the ${\cal O}(E^4)$ terms not restricted to
the self-interaction of gauge fields is obtained and
the one loop counter terms are determined explicitly in the $SU(N)$ model.
The size of the radiative correction to the phenomenological parameter $a$
is estimated to be about $10$\%.
We also obtain several matching conditions
to the conventional ChPT without the $\rho$ meson
in a renormalization scale independent manner.
Remarkably enough, the one loop effect of the $\pi$-$\rho$ system is shown
to imitate the tree level contribution\cite{kn:EGPR89}
from the $a_1$ meson in the Gasser--Leutwyler parameter $L_{10}$.
We find that the Gasser--Leutwyler estimates\cite{kn:GL84}
for $L_{9,10}$ are saturated by $\rho$ and its one loop contributions
without introducing non-minimal ${\cal O}(E^4)$ couplings in
$\pi$-$\rho$ system.

Let us start with a quick review of the BKUYY lagrangian\cite{kn:BKUYY}.
BKUYY decomposed the pion field $U=\exp(2iT^a \pi^a)$, transforming
$U\rightarrow g_L U g_R^\dagger$ under $SU(N)_L\times SU(N)_R$
into two parts,
\begin{equation}
  U = \xi_L^\dagger \xi_R.
\end{equation}
Arbitrariness in this decomposition can be regarded as a local symmetry
(hidden local symmetry, HLS):
\begin{equation}
  \xi_{L,R} \rightarrow h(x) \xi_{L,R}, \qquad h(x) \in SU(N)_{\rm HLS}.
\label{eq:transXi}
\end{equation}
This hidden local ``symmetry'' is redundant and
nothing to do with the $\pi$ system at this level.
By introducing a kinetic term for the gauge field associated with the HLS,
however,
it becomes a dynamical degree of freedom which may be identified as the
$\rho$ meson.

This idea leads us to an effective lagrangian for the $\pi$-$\rho$ system:
\begin{equation}
  {\cal L} = f^2 \tr\left( \tMCpb{\mu} \tMCpr{\mu} \right)
             + \frac{f^2}{4} \tr(\hat\chi + \hat\chi^\dagger)
             +a f^2 \tr\left( \tMClb{\mu} \tMClr{\mu} \right)
            -\frac{1}{2g^2} \tr(V_{\mu\nu} V^{\mu\nu}),
\label{eq:HLSlagr}
\end{equation}
where $\tMCpb{\mu}$ and $\tMClb{\mu}$ are defined by
\begin{equation}
\hat\alpha^{\mu}_{\perp \atop \parallel}
  \equiv
    -\frac{i}{2}\left[ (D^\mu \xi_L) \xi_L^\dagger \mp
                       (D^\mu \xi_R) \xi_R^\dagger
                \right].
\end{equation}
The vector field $V_\mu$ is identified with the $\rho$ meson
and we have introduced external
fields $\cL_\mu$, $\cR_\mu$, $s$ and $p$ in an analogous manner to
Gasser--Leutwyler\cite{kn:GL84}:
\begin{subequations}
\begin{eqnarray}
  D_\mu \xi_L &\equiv&  \partial_\mu \xi_L - iV_\mu \xi_L + i\xi_L \cL_\mu, \\
  D_\mu \xi_R &\equiv&  \partial_\mu \xi_R - iV_\mu \xi_R + i\xi_R \cR_\mu, \\
\hat\chi
  &\equiv& 2b \xi_L (s+ip) \xi_R^\dagger.
\end{eqnarray}
\end{subequations}
The constant $b$ measures the quark pair condensate in the chiral limit.
The first two terms then correspond to the conventional chiral lagrangian at
${\cal O}(E^2)$, while the last two terms are the
mass and the kinetic terms of the $\rho$ meson.
It should be stressed here that the $\rho$ meson acquires its mass
($m_\rho^2 = a g^2 f^2$) through the Higgs mechanism in the present model.

Now, we are ready to discuss a systematic expansion of the $\pi$-$\rho$
amplitudes based on the lagrangian \eq{eq:HLSlagr}.
We first study rules of counting the orders in the effective lagrangian.
Since the external vector fields ($\cL_\mu$, $\cR_\mu$)
couple to $\xi$ as gauge connections,
they should be balanced with a derivative in the order counting.
The $\rho$ meson field $V_\mu$ is assigned ${\cal O}(E)$ for the
same reason.
The order of $\hat\chi$ is determined as ${\cal O}(E^2)$,
since it produces the pion mass which should be balanced with $p^2$ in
the propagator.
It should be stressed that we need to assign ${\cal O}(E)$ to
the coupling $g$ in order for the kinetic term of $\rho$
to be ${\cal O}(E^2)$.
This assignment implies that the effective lagrangian is expanded
in terms of the mass of the vector meson.

We next construct the most general effective lagrangian with ${\cal O}(E^4)$.
It is convenient to make a list of operators having homogeneous
transformation in the HLS\@.
It is easy to show that
\begin{equation}
  \tMCpb{\mu}, \quad \tMClb{\mu}
\label{eq:MCform}
\end{equation}
are the only operators in this class at the lowest order (${\cal O}(E)$).
In addition to the covariant derivatives of \EQ{eq:MCform},
there exist operators in higher orders, e.g.,
\begin{equation}
 V_{\mu\nu}, \quad \hat\cV_{\mu\nu}, \quad \hat\cA_{\mu\nu}, \quad \hat\chi,
\end{equation}
where $V_{\mu\nu}$ is the field strength of the $\rho$ field and
$\hat\cV_{\mu\nu}$ and $\hat\cA_{\mu\nu}$ are defined as
\begin{subequations}
\begin{eqnarray}
  \hat\cV_{\mu\nu}
    &\equiv& \frac{1}{2}\left[
       \xi_R \cR_{\mu\nu} \xi_R^\dagger + \xi_L \cL_{\mu\nu} \xi_L^\dagger
     \right], \\
  \hat\cA_{\mu\nu}
    &\equiv& \frac{1}{2}\left[
       \xi_R \cR_{\mu\nu} \xi_R^\dagger - \xi_L \cL_{\mu\nu} \xi_L^\dagger
     \right],
\end{eqnarray}
\end{subequations}
with $\cR_{\mu\nu}$ and $\cL_{\mu\nu}$
being the field strength of $\cR_\mu$ and $\cL_\mu$, respectively.

The most general lagrangian at ${\cal O}(E^4)$
can now be constructed by taking a trace
of products of those operators and the covariant derivatives of
$\hat \alpha_{\mu\perp,\parallel}$.
It should be, however, noticed that the anti-symmetric combination
of the covariant derivatives of $\hat \alpha_{\mu\perp,\parallel}$ can be
expressed in terms of other operators:
\begin{subequations}
\begin{eqnarray}
 D_\mu \tMCpb{\nu} - D_\nu \tMCpb{\mu}
  &=& i[\tMClb{\mu}, \tMCpb{\nu}]+i[\tMCpb{\mu}, \tMClb{\nu}]
     -\hat\cA_{\mu\nu},
\label{eq:identity1}
  \\
 D_\mu \tMClb{\nu} - D_\nu \tMClb{\mu}
  &=& i[\tMClb{\mu}, \tMClb{\nu}]+i[\tMCpb{\mu}, \tMCpb{\nu}]
     +\hat\cV_{\mu\nu} - V_{\mu\nu}.
\label{eq:identity2}
\end{eqnarray}
\end{subequations}
Thus, it is sufficient to consider the symmetric combinations:
\begin{displaymath}
  D_\mu \hat \alpha_{\nu\perp,\parallel}
         + D_\nu \hat \alpha_{\mu\perp,\parallel}
        - \frac{1}{2} g_{\mu\nu} D_\rho \hat \alpha^{\rho}_{\perp,\parallel},
  \qquad
  D_\mu \hat \alpha^\mu_{\perp,\parallel}.
\end{displaymath}

The independent terms in ${\cal O}(E^4)$ lagrangian can be further reduced by
using the equations of motion:
\begin{subequations} \label{eq:onshell}
\begin{eqnarray}
  D_\mu \tMCpr{\mu}
    &=& i(1-a)[\tMClb{\mu},\tMCpr{\mu}]
        + i({\hat\chi}-{\hat\chi}^\dagger) + {\cal O}(E^4),
  \\
  D_\mu \tMClr{\mu}
    &=& {\cal O}(E^4),
  \\
  D_\mu V^{\mu\nu}
    &=& g^2 a f^2 \tMClr{\nu} + {\cal O}(E^4).
\end{eqnarray}
\end{subequations}

After a short algebra, we find the general form of
${\cal O}(E^4)$ lagrangian with the even intrinsic parity:
\begin{equation}
  {\cal L}_4 = {\cal L}_4^w + {\cal L}_4^x + {\cal L}_4^y + {\cal L}_4^z,
\end{equation}
where
\begin{subequations}
\begin{eqnarray*}
{\cal L}_4^w
  &=& w_1 \tr(\tMCpr{\mu}\tMCpb{\mu}(\hat\chi +\hat\chi^\dagger))
     +w_2 \tr(\tMCpr{\mu}\tMCpb{\mu})\tr(\hat\chi +\hat\chi^\dagger)
  \\
  & &+w_3 \tr(\tMClr{\mu}\tMClb{\mu}(\hat\chi +\hat\chi^\dagger))
     +w_4 \tr(\tMClr{\mu}\tMClb{\mu})\tr(\hat\chi +\hat\chi^\dagger)
     +w_5 \tr([\tMCpr{\mu}, \tMClb{\mu}](\hat\chi-\hat\chi^\dagger))
  \\
  & &+w_6 \tr((\hat\chi+\hat\chi^\dagger)^2)
     +w_7 (\tr(\hat\chi+\hat\chi^\dagger))^2
  \\
  & &+w_8 \tr((\hat\chi-\hat\chi^\dagger)^2)
     +w_9 (\tr(\hat\chi-\hat\chi^\dagger))^2
  \yesnumber\\
{\cal L}_4^x
  &=& x_1 m_\rho^2 \tr(\tMCpb{\mu}\tMCpr{\mu})
     +x_2 m_\rho^2 \tr(\tMClb{\mu}\tMClr{\mu})
     +x_3 \tr(V_{\mu\nu}V^{\mu\nu})
     +x_4 m_\rho^2 \tr(\hat\chi+\hat\chi^\dagger),
  \yesnumber\\
{\cal L}_4^y
  &=& y_1 \tr((\tMCpb{\mu}\tMCpr{\mu})^2)
     +y_2 \tr(\tMCpb{\mu}\tMCpb{\nu}\tMCpr{\mu}\tMCpr{\nu})
     +y_3 \tr((\tMClb{\mu}\tMClr{\mu})^2)
     +y_4 \tr(\tMClb{\mu}\tMClb{\nu}\tMClr{\mu}\tMClr{\nu})
  \\
  & &+y_5 \tr(\tMCpb{\mu}\tMCpr{\mu}\tMClb{\nu}\tMClr{\nu})
     +y_6 \tr(\tMCpb{\mu}\tMCpb{\nu}\tMClr{\mu}\tMClr{\nu})
  \\
  & &+y_7 \tr(\tMCpb{\mu}\tMCpb{\nu}\tMClr{\nu}\tMClr{\mu})
     +y_8 [ \tr(\tMCpb{\mu}\tMClr{\mu}\tMCpb{\nu}\tMClr{\nu})
           +\tr(\tMClb{\mu}\tMCpr{\mu}\tMClb{\nu}\tMCpr{\nu})]
  \\
  & &+y_9 \tr(\tMCpb{\mu}\tMClb{\nu}\tMCpr{\mu}\tMClr{\nu})
  \\
  & &+y_{10} (\tr(\tMCpb{\mu}\tMCpr{\mu}))^2
     +y_{11} \tr(\tMCpb{\mu}\tMCpb{\nu}) \tr(\tMCpr{\mu}\tMCpr{\nu})
     +y_{12} (\tr(\tMClb{\mu}\tMClr{\mu}))^2
  \\
  & &+y_{13} \tr(\tMClb{\mu}\tMClb{\nu}) \tr(\tMClr{\mu}\tMClr{\nu})
     +y_{14} \tr(\tMCpb{\mu}\tMCpr{\mu}) \tr(\tMClb{\nu}\tMClr{\nu})
     +y_{15} \tr(\tMCpb{\mu}\tMCpb{\nu}) \tr(\tMClr{\mu}\tMClr{\nu})
  \\
  & &+y_{16} (\tr(\tMCpb{\mu}\tMClr{\mu}))^2
     +y_{17} \tr(\tMCpb{\mu}\tMClb{\nu}) \tr(\tMCpr{\mu}\tMClr{\nu})
     +y_{18} \tr(\tMCpb{\mu}\tMClb{\nu}) \tr(\tMClr{\mu}\tMCpr{\nu}),
  \yesnumber\\
{\cal L}_4^z
  &=& z_1\tr(\hat\cV_{\mu\nu}\hat\cV^{\mu\nu})
     +z_2\tr(\hat\cA_{\mu\nu}\hat\cA^{\mu\nu})
     +z_3\tr(\hat\cV_{\mu\nu} V^{\mu\nu})
  \\
  & &+z_4 i\tr(V_{\mu\nu}\tMCpr{\mu}\tMCpr{\nu})
     +z_5 i\tr(V_{\mu\nu}\tMClr{\mu}\tMClr{\nu})
     +z_6 i\tr(\hat\cV_{\mu\nu}\tMCpr{\mu}\tMCpr{\nu})
     +z_7 i\tr(\hat\cV_{\mu\nu}\tMClr{\mu}\tMClr{\nu})
  \\
  & &+z_8
i\tr(\hat\cA_{\mu\nu}(\tMCpr{\mu}\tMClr{\nu}+\tMClr{\mu}\tMCpr{\nu})).
  \yesnumber
  \label{eq:fourth}
\end{eqnarray*}
\end{subequations}

The one loop calculations are required to make use of systematic
expansion of this kind\cite{kn:We79b}.
We adopt the background field technique\cite{kn:GL84} to avoid an
unnecessary complication of defining off-shell effective fields:
Naive calculation without a background field requires redefinitions
of the off-shell effective fields to maintain manifest chiral symmetry at
the loop level\cite{kn:AB81}.
Actually, the ``low energy theorem'' of the HLS\cite{kn:BKY85}
cannot be proven without an appropriate field redefinition in the
naive covariant  gauge calculation\cite{kn:KHY93}.
On the other hand,
the background field method maintains manifest symmetry at each step of
calculations as we will show in the following.

We introduce background fields $\bar\xi_L$, $\bar\xi_R$ and $\bar V_\mu$:
\begin{equation}
  \xi_L = \xi_S \xi_P^\dagger \bar \xi_L,  \qquad
  \xi_R = \xi_S \xi_P \bar \xi_R,          \qquad
  V_\mu = \bar V_\mu + v_\mu,
\end{equation}
where dynamical degrees of freedom are
denoted by $\xi_S$, $\xi_P$ and $v_\mu$:
\begin{equation}
  \xi_S = \exp \left( \frac{i u_S^a T^a}{\sqrt{a}f} \right), \qquad
  \xi_P = \exp \left( \frac{i u_P^a T^a}{f} \right),         \qquad
  v_\mu = g v_\mu^a T^a.
\end{equation}
Transformation properties under HLS  are
\begin{subequations} \label{eq:HLS2}
\begin{eqnarray}
  \bar \xi_{L,R} &\rightarrow&
    h(x) \bar \xi_{L,R}, \\
  \bar V_\mu     &\rightarrow&
    h(x) \bar V_\mu h^\dagger(x) +ih(x)\partial_\mu h^\dagger(x), \\
  u_{S,P}^a T^a &\rightarrow&
    h(x) u_{S,P}^a T^a h^\dagger(x), \\
  v_\mu^a T^a   &\rightarrow&
    h(x)v_\mu^a T^a h^\dagger(x).
\end{eqnarray}
\end{subequations}
Note here that the dynamical fields $u_P^a$, $u_S^a$ and $v_\mu^a$
are transformed linearly. Thus,
expansion of the chiral lagrangian in terms of these fields
does not violate the HLS\@.

In this formalism, a gauge fixing term for $v_\mu^a$ can
be introduced without violating the HLS of the background field $\bar V_\mu$
(background gauge\cite{kn:Ab82,kn:Ka82})
\begin{equation}
  {\cal L}_{\rm GF} = -\frac{1}{2}(\bar D^{ab\mu} v_\mu^b + m_\rho u_S^a)^2,
\label{eq:GFterm}
\end{equation}
with $\bar D_\mu$ being the ``covariant derivative'' on the background field:
\begin{equation}
  \bar D^{ab\mu} v_\mu^b T^a \equiv
  \partial^\mu v_\mu^a T^a -i [\bar V^\mu, v_\mu^a T^a].
\end{equation}
The Faddeev--Popov determinant associated with the gauge fixing term
\eq{eq:GFterm} is
\begin{equation}
  {\cal L}_{\rm FP} = i \bar c^a \left( \bar D^{ab\mu} \bar D^{bc}_\mu
                                       + m_\rho^2 \delta^{ac} \right) c^c
                     + \cdots,
\label{eq:FP}
\end{equation}
where $\cdots$ stands for the interaction terms of the
dynamical fields $u_S^a$, $v_\mu^a$  and the FP ghosts.

Now, it is straightforward to evaluate the one loop contribution
to the above ${\cal O}(E^4)$ lagrangian coefficients.
After tedious manipulation in the heat kernel method\cite{kn:GL84},
we obtain coefficients of the logarithmic divergences
in the dimensional regularization scheme as\footnote{
Renormalization scale independent quadratic divergences can be
renormalized by the redefinition of parameters in the
${\cal O}(E^2)$ lagrangian.}
\begin{equation}
  w_i = w_i^r - \frac{\Gamma_{wi}}{4}\left[
            \frac{\Gamma(2-d/2)}{(4\pi)^{d/2}} + \frac{1}{(4\pi)^2}
          \right],
\end{equation}
and similarly for the coefficients $x_i$, $y_i$ and $z_i$.
The result is summarized in table 1.
These coefficients determine the running of the renormalized parameters,
\begin{equation}
  \mu \frac{\partial}{\partial \mu} w_i^r(\mu)
      = -\frac{\Gamma_{wi}}{2(4\pi)^2}, \quad \cdots,
\label{eq:RGE}
\end{equation}
which can be considered as a measure of the effect of radiative corrections.

It should be noted, however,
these renormalization group equations \eq{eq:RGE} are derived in the mass
independent scheme and thus the scale $\mu$ is not directly related to
the actual physics scale,
especially at the low energy region below $m_\rho$
where the dynamical degree of freedom of the $\rho$ meson is frozen out.
Actually, as we will show in the following,
finite part of the one loop integral is important to make the
matching conditions with the conventional ChPT without the $\rho$ meson.
To construct effective couplings valid in the whole energy region
we find it convenient to introduce the *-functions in
analogous way with those defined
in the electroweak effective lagrangian\cite{kn:KL89}.

The *-functions are functions of momentum and independent of
renormalization schemes.
They are designed to reproduce the real part of the scattering
amplitude at the one loop level by
replacing the tree level parameters.
Such functions can be easily calculated once the effective action
is determined.
For example, the *-functions corresponding to $f$, $a$ and $g$ are calculated
by extracting the corresponding terms from the real part of the effective
action:
\begin{eqnarray*}
  \Re \Gamma
  &=& \int d^4x d^4z \tilde f_*^2(z)
       \tr(\tMCpb{\mu}(x+z) \tMCpr{\mu}(x))
  \\
  & & +\int d^4x d^4z \tilde a_*(z)
       f^2 \tr(\tMClb{\mu}(x+z) \tMClr{\mu}(x))
  \\
  & & -\int d^4x d^4z
       \frac{1}{2 \tilde g_*^2(z)} \tr(V_{\mu\nu}(x+z) V^{\mu\nu}(x))
      \cdots,
  \yesnumber
\end{eqnarray*}
with $\tilde f_*^2$, $\tilde a_*$ and $1/\tilde g_*^2$ being Fourier
transform of *-functions $f_*^2(p^2)$, $a_*(p^2)$ and $1/g_*^2(p^2)$,
respectively.
These *-functions can be determined from the two-point functions of
the building blocks $\tMCpb{\mu}$, $\tMClb{\mu}$,\dots.
In the chiral limit, we find:
\begin{subequations}\label{eq:HLSstar}
\begin{eqnarray}
  f^2_*(p^2)
    &=& f^2 + m_\rho^2 \left\{
         x_1^r(m_\rho) + \frac{Na}{(4\pi)^2} \left[
           1+B_0(1,0; \frac{-p^2}{m_\rho^2})
            -\frac{1}{4} B_2(1,0; \frac{-p^2}{m_\rho^2})
         \right] \right\},
\label{eq:star_f}
    \\
  a_*(p^2) f^2
    &=& a f^2 + m_\rho^2 \left\{
         x_2^r(m_\rho) + \frac{N}{(4\pi)^2} \left[
           1 + B_0(1,1; \frac{-p^2}{m_\rho^2})
         \right]\right\},
\label{eq:star_a}
    \\
  \frac{1}{g_*^2(p^2)}
  &=& \frac{1}{g^2} -2 x_3^r(m_\rho)
  \nonumber\\
  & & +\frac{N}{(4\pi)^2} \left[
        -\frac{a^2}{24}\Re\ln\left(\frac{-p^2}{m_\rho^2}\right)
        +\frac{31}{24}+\frac{5a^2}{72}
        +4 B_0(1,1; \frac{-p^2}{m_\rho^2})
        -\frac{9}{8} B_4(1,1; \frac{-p^2}{m_\rho^2})
       \right],
  \nonumber\\
  & &
\end{eqnarray}
\end{subequations}
with $B_0$, $B_2$ and $B_4$ being defined by:
\begin{eqnarray*}
  B_0(X,Y; P) &\equiv& \Re \int_0^1 d\alpha
                       \ln(\alpha X + (1-\alpha) Y + \alpha(1-\alpha) P), \\
  B_2(X,Y; P) &\equiv& \Re \int_0^1 d\alpha (1-2\alpha)
                       \ln(\alpha X + (1-\alpha) Y + \alpha(1-\alpha) P), \\
  B_4(X,Y; P) &\equiv& \Re \int_0^1 d\alpha (1-2\alpha)^2
                       \ln(\alpha X + (1-\alpha) Y + \alpha(1-\alpha) P).
\end{eqnarray*}
The typical size of ${\cal O}(E^4)$ corrections can be estimated in
this calculation.
Expanding \EQ{eq:star_a} around $p^2=0$, we obtain
\begin{displaymath}
  a_*(p^2)f^2-a_*(0)f^2
  = -\frac{1}{6}\frac{N}{(4\pi)^2} p^2
      + {\cal O}\left(\frac{p^4}{m_\rho^2}\right)
\end{displaymath}
and find $a_*(p^2)$ is insensitive to the loop effect.
Actually, $a_*(p^2)$ receives only $10$\% level radiative correction:
$a_*(m_\rho^2) = a_*(0) -0.17$ in $SU(2)$ model for $f=88$MeV.

We can extract information of ${\cal O}(E^4)$ parameters through the matching
conditions with the conventional ChPT without the $\rho$ meson which is
already studied extensively by Gasser and Leutwyler\cite{kn:GL84}.
Gasser--Leutwyler lagrangian is given by
\begin{eqnarray*}
 {\cal L}_{\rm GL}
  &=& \frac{F^2}{4} \tr((D_\mu U)^\dagger (D^\mu U))
    + \frac{F^2}{4} \tr(U^\dagger X + X^\dagger U)
  \\
  & & + L_1 (\tr((D_\mu U)^\dagger (D^\mu U)))^2 +\cdots \\
  & & -i L_9 \tr ( \cR_{\mu\nu} (D^\mu U)^\dagger (D^\nu U)
                  +\cL_{\mu\nu} (D^\mu U) (D^\nu U)^\dagger ) \\
  & & + L_{10} \tr( U^\dagger \cL_{\mu\nu} U \cR^{\mu\nu} )
      + H_1 \tr(\cL_{\mu\nu} \cL^{\mu\nu}+\cR_{\mu\nu} \cR^{\mu\nu})
      + H_2 \tr(X^\dagger X),
\yesnumber
\end{eqnarray*}
with $X=2B(s+ip)$ and $F$ being the NG boson decay constant in the
chiral limit.
In the following, we focus our attention on the parameters $L_{9,10}$.
It is convenient to introduce the *-functions corresponding to
the parameters $L_{9,10}$ and $H_1$.
In the chiral limit, we find
\begin{subequations}
\begin{eqnarray}
  F_*^2(p^2)
    &=& F^2, \\
  L_9^*(p^2)
    &=& L_9^r -\frac{N}{24(4\pi)^2}
               \left(-\frac{5}{3}+\Re\ln\frac{-p^2}{\mu^2}\right),
    \\
  L_{10}^*(p^2)
    &=& L_{10}^r +\frac{N}{24(4\pi)^2}
               \left(-\frac{5}{3}+\Re\ln\frac{-p^2}{\mu^2}\right),
    \\
  H_1^*(p^2)
    &=& H_1^r +\frac{N}{48(4\pi)^2}
               \left(-\frac{5}{3}+\Re\ln\frac{-p^2}{\mu^2}\right).
\end{eqnarray}
\end{subequations}

To obtain the matching conditions, we integrate out $\tMClb{\mu}$ and
$V_{\mu\nu}$ from the effective action by making use of
the equations of motion:
\begin{subequations}
\begin{equation}
  \tMClb{\mu} = \frac{1}{m_\rho^2} {\cal O}(E^3),
\end{equation}
and
\begin{equation}
  V_{\mu\nu} = \hat\cV_{\mu\nu} + i[\tMCpb{\mu},\tMCpb{\nu}]
              + \frac{1}{m_\rho^2} {\cal O}(E^4).
\end{equation}
\end{subequations}
The latter equation can be derived from
\begin{displaymath}
  D_\mu\tMClb{\nu} - D_\nu\tMClb{\mu} -i[\tMClb{\mu}, \tMClb{\nu}]
  =\frac{1}{m_\rho^2} {\cal O}(E^4),
\end{displaymath}
and \EQ{eq:identity2}.
Thus, we obtain:
\begin{subequations} \label{eq:Matching}
\begin{eqnarray}
  F_*^2(p^2)
    &=& f_*^2(p^2) + {\cal O} \left(\frac{p^2}{m_\rho^2}\right), \\
  -8L_9^*(p^2)
    &=& -\frac{2}{g_*^2(p^2)} + 2z_3^*(p^2) + z_4^*(p^2, -p^2, 0)
    \nonumber\\
    & & \qquad
        + z_6^*(p^2, -p^2,0 ) + {\cal O} \left(\frac{p^2}{m_\rho^2}\right), \\
  L_{10}^*(p^2)  + 2H_1^*(p^2)
    &=& -\frac{1}{2g_*^2(p^2)} + z_1^*(p^2) + z_3^*(p^2)
         + {\cal O} \left(\frac{p^2}{m_\rho^2}\right), \\
 -L_{10}^*(p^2)  + 2H_1^*(p^2)
    &=& z_2^*(p^2) + \frac{1}{2} f_*^2{}'(0)
                   + {\cal O} \left(\frac{p^2}{m_\rho^2}\right).
\end{eqnarray}
\end{subequations}

The *-functions corresponding to $z_1$, $z_2$ and $z_3$
are determined as
\begin{subequations}
\begin{eqnarray}
z_1^*(p^2)
  &=& z_1^r(m_\rho) + \frac{N}{16(4\pi)^2}\left[
        \frac{(2-a)^2}{3}\Re\ln\left(\frac{-p^2}{m_\rho^2}\right)
        -\frac{5}{9}(2-a)^2 + \frac{1}{3} + B_4(1,1; \frac{-p^2}{m_\rho^2})
      \right],
  \nonumber\\
  & &
  \\
z_2^*(p^2)
  &=& z_2^r(m_\rho) + \frac{Na}{8(4\pi)^2} \left[
        \frac{1}{3} + B_4(1,0; \frac{-p^2}{m_\rho^2})
      \right],
  \\
z_3^*(p^2)
  &=& z_3^r(m_\rho) + \frac{N}{8(4\pi)^2}\left[
        \frac{a (2-a)}{3} \Re \ln \left(\frac{-p^2}{m_\rho^2}\right)
        -\frac{5}{9} a(2-a) + \frac{1}{3} + B_4(1,1; \frac{-p^2}{m_\rho^2})
      \right].
  \nonumber\\
  & &
\end{eqnarray}

The *-functions corresponding to $z_4$ and $z_6$ depend not simply on
single momentum, since they lead to non-oblique three point
vertices of $\rho$-$\pi$-$\pi$ and photon-$\pi$-$\pi$, respectively:
\begin{displaymath}
  z_4^*(p^2,q^2,p\cdot q), \qquad z_6^*(p^2, q^2, p\cdot q),
\end{displaymath}
where $p$ and $q$ stand for incoming momentum of $\rho$ (photon) and
the difference of two incoming pion momenta, respectively.
These *-functions become simple when we take $q^2 = -p^2$, $p\cdot q = 0$
(corresponding to the on-shell of pions in the 3-point vertex):
\begin{eqnarray}
z_4^*(p^2, -p^2, 0)
  &=& z_4^r(m_\rho) + \frac{N}{4(4\pi)^2} \left[
        \frac{a(2-a)}{3} \Re\ln\left(\frac{-p^2}{m_\rho^2}\right)
       -\frac{5}{9} a(2-a) + \frac{1}{3}
      \right. \nonumber\\
  & & \left.
       + B_4(1,1; \frac{-p^2}{m_\rho^2})
       -a C(1,0; \frac{-p^2}{m_\rho^2}) -a^2 C(0,1; \frac{-p^2}{m_\rho^2})
      \right],
  \\
z_6^*(p^2, -p^2, 0)
  &=& z_6^r(m_\rho) + \frac{N}{4(4\pi)^2} \left[
        \frac{(2-a)^2}{3} \Re\ln\left(\frac{-p^2}{m_\rho^2}\right)
       -\frac{5}{9} (2-a)^2 + \frac{1}{3}
      \right. \nonumber\\
  & & \left.
       + B_4(1,1; \frac{-p^2}{m_\rho^2})
       -a C(1,0; \frac{-p^2}{m_\rho^2}) -a(2-a) C(0,1; \frac{-p^2}{m_\rho^2})
      \right],
\end{eqnarray}
with $C$ being defined by
\begin{displaymath}
  C(X,Y; P) \equiv \frac{1}{6}
                   + \Re\int_0^1 d\alpha \int_0^1 d\beta \alpha(2\alpha-1)
                     \ln(\alpha X + (1-\alpha) Y + \alpha^2 \beta(1-\beta) P).
\end{displaymath}
\end{subequations}
It should be stressed here that the infrared singularity
($\ln (p^2)$) in $z_i^*$ ($i=1,2,3,4,6$) does not appear
at $a=2$ (vector dominance).

Note here that the coefficients of $\ln(p^2)$ arise from the finite part of
the one loop integral and agree exactly
in both sides of \EQ{eq:Matching}.
It should also be stressed that the matching conditions \eq{eq:Matching}
are free from the renormalization scale ambiguity in contrast with
the previous tree level analysis of the resonance effect to Gasser--Leutwyler
parameters\cite{kn:EGPR89}.

By using the values of $SU(2)$ Gasser--Leutwyler
parameters\cite{kn:GL84,kn:EGPR89}
$L_9^r(550\mbox{MeV}) = (7.2\pm0.7) \cdot 10^{-3}$,
$L_{10}^r(550\mbox{MeV}) = (-5.5\pm0.3) \cdot 10^{-3}$
and $f=f_*(0)=88\mbox{MeV}$, $m_\rho=757\mbox{MeV}$ estimated in the
chiral limit,
we find a cancellation among $L_{9,10}^*$, $1/g_*^2$ and $f_*^2{}'$
in \eq{eq:Matching} and obtain:
\begin{subequations} \label{eq:result}
\begin{eqnarray}
  \frac{1}{2}(z_1^*(0) - z_2^*(0) + z_3^*(0))
    &=& (-0.6 \pm 0.3) \cdot 10^{-3}, \\
  \frac{1}{8}(2 z_3^*(0) + z_4^*(0,0,0) + z_6^*(0,0,0) )
    &=& (0.7 \pm 0.7) \cdot 10^{-3},
\end{eqnarray}
\end{subequations}
or
\begin{subequations} \label{eq:result2}
\begin{eqnarray}
  \frac{1}{2}(z_1^*(m_\rho^2) - z_2^*(m_\rho^2) + z_3^*(m_\rho^2))
    &=& (0.1 \pm 0.3) \cdot 10^{-3}, \\
  \frac{1}{8}(2 z_3^*(m_\rho^2) + z_4^*(m_\rho^2,-m_\rho^2,0)
                                + z_6^*(m_\rho^2,-m_\rho^2,0) )
    &=& (0.8 \pm 0.7) \cdot 10^{-3},
\end{eqnarray}
\end{subequations}
where we have assumed the vector dominance $a=a_*(0)=2$ and used
$$
  g_*^2(m_\rho^2) = \frac{m_\rho^2}{a_*(m_\rho^2) f^2}.
$$
There is no strong indication for the presence of the non-minimal couplings
$z_i^*$.
In other words, the Gasser--Leutwyler estimates for
$L_{9,10}$ are saturated by the $\rho$ meson and its one loop contribution
without introducing non-minimal ${\cal O}(E^4)$ terms.
This result contrasts with the tree level analysis indicating
that $L_{10}$ is not saturated by the $\rho$ meson contribution alone
but saturated by the combined effect of the $\rho$ meson and
the $a_1$ meson\cite{kn:EGPR89}.
Actually, the tree level contribution from the $a_1$ resonance is imitated by
the one loop effect $f_*^{2}{}'$ in the matching conditions \eq{eq:Matching}.

In this paper, we have formulated the ChPT of the $\pi$-$\rho$ system based on
the effective lagrangian\cite{kn:BKUYY} with the hidden local symmetry.
The size of the radiative correction to the phenomenological parameter $a$
is estimated to be about $10$\%.
We also obtain the matching conditions to the conventional ChPT of
Gasser--Leutwyler in the renormalization scale independent manner.
It should be emphasized that
the one loop effect imitates the tree level contribution of
the $a_1$ resonance in the matching condition for $L_{10}$.
We find that Gasser--Leutwyler's estimates for
$L_{9,10}$ are saturated by the $\rho$ meson and its one loop contribution
without introducing the ${\cal O}(E^4)$ couplings.

Technical details of the present work will be published elsewhere.

The author thanks K. Yamawaki for introducing him to the
importance of this subject.
He is very grateful to Y. Okada
for continuous encouragements and fruitful discussions.
He thanks also K. Hagiwara,  M. Harada, M. Kobayashi, T. Kugo,
S. Matsumoto, V.A. Miransky, A.I. Sanda, M. Tanaka and T.N. Truong
for enlightening discussions.

\newpage
\begin{table}
\begin{center}
\begin{tabular}{|c||c|c|c|c|} \hline
       & \hspace{2.5cm} $\Gamma_{wi}$  \hspace{2.5cm}
       & \hspace{0.5cm} $a=1$ \hspace{0.5cm}
       & \hspace{0.5cm}$a=2$ \hspace{0.5cm}\\
  \hline
  $w_1$   & $2N(4-3a)$ & $2N$ & $-4N$ \\
  $w_2$   & $2(4-3a)$ & $2$ & $-4$ \\
  $w_3$   & $2Na^2$ & $2N$ & $8N$ \\
  $w_4$   & $2a^2$ & $2$ & $8$ \\
  $w_5$
        & $\frac{3}{2} Na(1-a)$ & $0$ & $-3N$ \\
  $w_6$
        & $\frac{2}{N} (N^2-4)$ & $\frac{2}{N}(N^2-4)$ &$\frac{2}{N} (N^2-4)$
\\
  $w_7$
        & $\frac{2}{N^2} (N^2+2)$ & $\frac{2}{N^2}(N^2+2)$
        & $\frac{2}{N^2} (N^2+2)$ \\
  $w_8$ & $-Na$ & $-N$ & $-2N$ \\
  $w_9$ & $a$ & $1$ & $2$ \\
  \hline
\end{tabular}
\begin{tabular}{|c||c|c|c|c|} \hline
       & \hspace{2.5cm} $\Gamma_{xi}$  \hspace{2.5cm}
       & \hspace{0.5cm} $a=1$ \hspace{0.5cm}
       & \hspace{0.5cm} $a=2$ \hspace{0.5cm} \\
  \hline
  $x_1$   & $-3Na$ & $-3N$ & $-6N$ \\
  $x_2$   & $-3N$  & $-3N$ & $-3N$ \\
  $x_3$   & $\frac{N}{12}(87-a^2)$ & $\frac{43N}{6}$ & $\frac{83N}{12}$ \\
  $x_4$   & $0$    & $0$   & $0$ \\
  \hline
\end{tabular}
\begin{tabular}{|c||c|c|c|c|} \hline
       & \hspace{2.5cm} $\Gamma_{zi}$  \hspace{2.5cm}
       & \hspace{0.5cm} $a=1$ \hspace{0.5cm}
       & \hspace{0.5cm} $a=2$ \hspace{0.5cm} \\
  \hline
  $z_1$   & $-\frac{N}{12}(5-4a+a^2)$ & $-\frac{N}{6}$ & $-\frac{N}{12}$ \\
  $z_2$   & $-\frac{N}{6}a$ & $-\frac{N}{6}$ & $-\frac{N}{3}$ \\
  $z_3$   & $-\frac{N}{6}(1+2a-a^2)$   & $-\frac{N}{3}$ & $-\frac{N}{6}$ \\
  $z_4$   & $-\frac{N}{6}(2+3a-3a^2)$  & $-\frac{N}{3}$ & $\frac{2N}{3}$ \\
  $z_5$   & $-\frac{N}{6}(1+2a^2-a^3)$ & $-\frac{N}{3}$ & $-\frac{N}{6}$ \\
  $z_6$   & $-\frac{N}{6}(2-a)(5-3a)$  & $-\frac{N}{3}$ & $0$ \\
  $z_7$   & $-\frac{N}{6}(1+4a-4a^2+a^3)$ & $-\frac{N}{3}$ & $-\frac{N}{6}$ \\
  $z_8$   & $\frac{N}{6}a(1+a)$ & $\frac{N}{3}$  & $N$ \\
\hline
\end{tabular}
\end{center}
\end{table}
\newpage
\begin{table}
\begin{center}
\begin{tabular}{|c||c|c|c|c|} \hline
       & \hspace{2.5cm} $\Gamma_{yi}$  \hspace{2.5cm}
       & \hspace{0.5cm} $a=1$ \hspace{0.5cm}
       & \hspace{0.5cm} $a=2$ \hspace{0.5cm} \\
  \hline
  $y_1$   & $\frac{N}{6}(7-11a+5a^2)$ & $\frac{N}{6}$ & $\frac{5N}{6}$ \\
  $y_2$   & $\frac{N}{12}(10-14a+5a^2)$ & $\frac{N}{12}$ & $\frac{N}{6}$ \\
  $y_3$   & $\frac{N}{12}(1-2a^2+2a^3+a^4)$ & $\frac{N}{6}$ & $\frac{25N}{12}$
\\
  $y_4$   & $\frac{N}{24}(1+4a^2-4a^3+a^4)$ & $\frac{N}{12}$ & $\frac{N}{24}$
\\
  $y_5$   & $\frac{N}{6}a(1+6a-5a^2)$ & $\frac{N}{3}$ & $-\frac{7N}{3}$ \\
  $y_6$   & $\frac{N}{6}(1+4a-5a^2+2a^3)$ & $\frac{N}{3}$ & $\frac{5N}{6}$ \\
  $y_7$   & $\frac{N}{6}(-1+10a-13a^2+6a^3)$ & $\frac{N}{3}$ & $\frac{5N}{2}$
 \\
  $y_8$   & $\frac{N}{6}a^2(4-3a)$ & $\frac{N}{6}$ & $-\frac{4N}{3}$ \\
  $y_9$   & $\frac{N}{6}a^2$ & $\frac{N}{6}$ & $\frac{2N}{3}$ \\
  $y_{10}$ & $\frac{1}{4}(8-12a+5a^2)$ & $\frac{1}{4}$ & $1$ \\
  $y_{11}$ & $\frac{1}{2}(8-12a+5a^2)$ & $\frac{1}{2}$ & $2$ \\
  $y_{12}$ & $\frac{1}{8}(1+a^4)$ & $\frac{1}{4}$ & $\frac{17}{8}$ \\
  $y_{13}$ & $\frac{1}{4}(1+a^4)$ & $\frac{1}{2}$ & $\frac{17}{4}$ \\
  $y_{14}$ & $\frac{1}{6}a(1+7a-5a^2)$ & $\frac{1}{2}$ & $-\frac{5}{3}$ \\
  $y_{15}$ & $\frac{1}{3}a(7-5a+a^2)$ & $1$ & $\frac{2}{3}$ \\
  $y_{16}$ & $\frac{1}{3}a(7-5a+a^2)$ & $1$ & $\frac{2}{3}$ \\
  $y_{17}$ & $\frac{1}{3}a(1+7a-5a^2)$ & $1$ & $-\frac{10}{3}$ \\
  $y_{18}$ & $\frac{1}{3}a(7-5a+a^2)$ & $1$ & $\frac{2}{3}$ \\
  \hline
\end{tabular}
\end{center}
\caption{Coefficients of counter terms}
\end{table}

\newpage
\begin{thebibliography}{99}
\bibitem{kn:We79b}
  S. Weinberg, Physica {\bf 96A} (1979) 327.
\bibitem{kn:GL84}
  J. Gasser and H. Leutwyler, Ann. Phys. (N.Y.) {\bf 158} (1984) 142;
  Nucl. Phys. {\bf B250} (1985) 465.
\bibitem{kn:HT90}
  B. Holdom and J. Terning, Phys. Lett. {\bf B247} (1990) 88;
  M. Golden and L. Randall, Nucl. Phys. {\bf B361} (1991) 3;
  H. Georgi, Nucl. Phys. {\bf B363} (1991) 301;
  A.F. Falk, M. Luke, E.H. Simmons,
  Nucl. Phys. {\bf B365} (1991) 523.
\bibitem{kn:BKUYY}
  M. Bando, T. Kugo, S. Uehara, K. Yamawaki and T. Yanagida,
  Phys. Rev. Lett. {\bf 54} (1985) 1215;
  For a review, see
  M. Bando, T. Kugo and K. Yamawaki, Phys. Reports {\bf 164} (1988) 218.
\bibitem{kn:We68}
  S. Weinberg, Phys. Rev. {\bf 166} (1968) 1568.
\bibitem{kn:SaTXT}
  J.J. Sakurai, {\it Currents and Mesons}\ (Univ. Chicago Press, Chicago,
1969).
\bibitem{kn:KS66}
  K. Kawarabayashi and M. Suzuki, Phys. Rev. Lett. {\bf 16} (1966) 255;
  Riazuddin and Fayyazuddin, Phys. Rev. {\bf 147} (1966) 1071.
\bibitem{kn:Ge89}
  H. Georgi, Phys. Rev. Lett. {\bf 63} (1989) 1917;
             Nucl. Phys. {\bf B331} (1990) 311.
\bibitem{kn:Ko92}
  P. Ko, Phys. Rev. {\bf D44} (1991) 139;
         Phys. Rev. {\bf D46} (1992) 3813;
  M. Finkemeier,
         Phys. Rev. {\bf D47} (1993) 1933.
\bibitem{kn:EGPR89}
  G. Ecker, J. Gasser, H. Leutwyler, A. Pich and E. de Rafael,
    Phys. Lett. {\bf B223} (1989) 425.
  See also
  G. Ecker, J. Gasser, A. Pich and E. de Rafael,
    Nucl. Phys. {\bf B321} (1989) 311;
  J.F. Donoghue, C. Ramirez and G. Valencia,
    Phys. Rev. {\bf D39} (1989) 1947.
\bibitem{kn:BESS}
  R. Casalbuoni, S. de Curtis, D. Dominici and R. Gatto,
  Phys. Lett. {\bf B155}, (1985) 95;
  Nucl. Phys. {\bf B282}, (1987) 235.
\bibitem{kn:CK91}
  G. Cveti\v{c} and R. K\"{o}gerler, Nucl. Phys. {\bf B363} (1991) 401.
\bibitem{kn:HY92}
  M. Harada and K. Yamawaki, Phys. Lett. {\bf B297} (1992) 151.
\bibitem{kn:BKY85}
  M. Bando, T. Kugo and K. Yamawaki, Nucl. Phys. {\bf B259} (1985) 493;
                                     Prog. Theor. Phys. {\bf 73} (1985) 1541.
\bibitem{kn:AB81}
  T. Appelquist and C. Bernard, Phys. Rev. {\bf D23} (1981) 425.
\bibitem{kn:KHY93}
  M. Harada, T. Kugo and K. Yamawaki, Nagoya preprint DPNU-93-01;
                                      Kyoto  preprint KUNS-1179 HE(TH) 93/02.
\bibitem{kn:Ab82}
  For a review, see
  L.F. Abbott, Acta. Phys. Pol. {\bf B13} (1982) 33.
\bibitem{kn:Ka82}
  Yu.N. Kafiev, Nucl. Phys. {\bf B201} (1982) 341.
\bibitem{kn:KL89}
  D.C. Kennedy and B.W. Lynn, Nucl. Phys. {\bf B322} (1989) 1.
\end{thebibliography}
\end{document}